\documentclass[aps,twocolumn,superscriptaddress,amsmath]{revtex4}

\pdfoutput=1

\usepackage{dcolumn}
\usepackage{bm}
\usepackage{amsfonts}
\usepackage[pdftex]{color,graphicx}
\usepackage{amssymb}
\usepackage{rotate}
\usepackage{subfigure}
\usepackage{subfigure}
\usepackage{amsmath}
\usepackage{amsthm}
\usepackage{hyperref}

\newcommand{\ket}[1]{\ensuremath{\left| #1 \right\rangle}}
\newcommand{\bra}[1]{\ensuremath{\left\langle #1 \right|}}

\newcommand{\cf}{{\it cf.~}}

\renewcommand{\-}{\,-\,}

\newcommand{\ZZ}{\mathbb{Z}}
\newcommand{\CC}{\mathbb{C}}

\newcommand{\CP}{\mathbb{CP}}

\newcommand{\set}[1]{\{ #1 \}}
					
\let\oldmarginpar\marginpar
\renewcommand\marginpar[1]{\-\oldmarginpar[\raggedleft\tiny #1]%
{\raggedright\tiny #1}}

\newtheorem{theorem}{Theorem}

\newtheorem{defn}{Definition}

\graphicspath{{figs/}}

\begin{document}


\title{On product, generic and random generic quantum satisfiability}
\hypersetup{pdftitle={On product, generic and random generic satisfiability},
	pdfauthor={C.R. Laumann, A.M. L\"auchli, R. Moessner, A. Scardicchio and S.L. Sondhi}}
	
\author{C. R. Laumann}
\affiliation{Department of Physics, Princeton University, Princeton, NJ 08544, USA}

\author{A. M. L\"auchli}
\affiliation{Max Planck Institut f¨ur Physik Komplexer Systeme, %
	01187 Dresden, Germany}

\author{R. Moessner}
\affiliation{Max Planck Institut f¨ur Physik Komplexer Systeme, %
	01187 Dresden, Germany}

\author{A. Scardicchio}
\affiliation{Abdus Salam International Centre for %
Theoretical Physics, Strada Costiera 11, 34014 Trieste, Italy}

\author{S. L. Sondhi}
\affiliation{Department of Physics, Princeton University, Princeton, NJ 08544}

\date{1 July 2010}


\begin{abstract}
	We report a cluster of results on $k$-QSAT, the problem of quantum satisfiability for $k$-qubit projectors which generalizes classical satisfiability with $k$-bit clauses to the quantum setting. First we define the NP-complete problem of product satisfiability and give a geometrical criterion for deciding when a QSAT interaction graph is product satisfiable with positive probability. We show that the same criterion suffices to establish quantum satisfiability for \emph{all} projectors. Second, we apply these results to the random graph ensemble with generic projectors and obtain improved lower bounds on the location of the SAT--unSAT transition. Third, we present numerical results on random, generic satisfiability which provide estimates for the location of the transition for $k=3$ and $k=4$ and mild evidence for the existence of a phase which is satisfiable by entangled states alone.
\end{abstract}

\maketitle

\tableofcontents


\section{Introduction} 
\label{sec:introduction}

The central achievement of classical complexity theory was the discovery that certain natural classes of constraint satisfaction problems, prototypically boolean satisfiability (SAT), can encode all verifiable decision problems in an efficient manner. This insight, encapsulated in the Cook-Levin theorem, and the general consensus that $P\ne NP$, imply that any classical approach to solving instances of SAT must in general take exponentially long \cite{Garey:1979ud}. Unfortunately, the recent development of an analogous set of quantum complexity theoretic results suggest that a similar situation holds even for quantum computers: in this case, quantum satisfiability (QSAT) can encode any problem in the class QMA$_1$%
\footnote{This is the class of decision problems which can be verified efficiently on a quantum computer with perfect completeness -- \emph{ie.} in which the verifier never rejects a valid proof. QMA$_1$ is closely related to the somewhat more general class QMA, which does not require perfect completeness.} %
and ought therefore resist solution by even the most subtle quantum attacks \cite{Bravyi:2006p4315}.

These hardness results, while rigorous, are in some sense unsatisfactory. They imply that (quantum) satisfiability is exponentially hard to solve in the worst-case, but shed little light on what separates the worst-case instance from the trivial. In this pursuit, it is useful to introduce a measure on instances of satisfiability and study the statistical features of the ensemble rather than the worst-case behavior of the whole class. This approach provides control parameters with which to tune our attention across various regimes of difficulty and also allows us to bring to bear the full arsenal of statistical mechanics and probability theory \cite{ScottKirkpatrick05271994,Monasson:1999p25}.

This paper is devoted to such a study in the context of {\em quantum} complexity theory. We study quantum satisfiability ($k$-QSAT), the decision problem of determining whether a given Hamiltonian $H$ acting on $N$ qubits has a ground state with precisely zero energy. 
The corresponding Hamiltonian has the form
\begin{equation}
	H = \sum_{m=1}^M \Pi^m
\end{equation}
where $\Pi^m$ is a $k$-local (\emph{i.e.} $k$-qubit) projector associated with the (hyper-)edge $m = (m_1, m_2,\cdots, m_k)$ of an interaction graph $G$. Such a Hamiltonian has a ground state $\ket{\psi}$ of zero energy if and only if $\ket{\psi}$ is simultaneously annihilated by all of the $\Pi^m$.  Furthermore, an instance of $k$-QSAT must come with a promise $\delta = 1 / \textrm{poly(N)}$ that the ground state energy of $H$, should it be greater than zero, will actually be greater than $\delta$. We will have little more to say regarding the promise gap in this paper, although we believe that it is satisfied in the thermodynamic limit of the ensembles we consider.

In Bravyi's original work introducing the $k$-QSAT problem, it was shown to be QMA$_1$-complete\cite{Bravyi:2006p4315} for $k\ge 4$. In a previous paper\cite{Laumann:2010fk}, we introduced a random ensemble for $k$-QSAT and worked out the coarse features of its phase diagram in the large $N$ limit.
There are two sources of randomness in the $k$-QSAT ensemble: 1. the discrete choice of interaction graph $G$ and 2. the continuous choice of projectors $\Pi^m$ associated to each edge. The latter choice is particularly powerful: \emph{generic} choices of projectors reduce quantum satisfiability to a graph, rather than Hamiltonian, property. 
More precisely, for fixed $G$, the dimension $R_G = |\ker H|$ of the satisfying subspace of $H$ is \emph{almost always} minimal with respect to the choice of projectors $\Pi^m$. Thus, if $G$ can be frustrated by \emph{some} choice of projectors, it is frustrated for almost all such choices. This ``geometrization'' property allows us to make strong statements about the quantum satisfiability of Hamiltonians associated with both random and non-random graphs and even non-generic choices of projector, both analytically and numerically. Throughout this paper, we use the term \emph{generic} to refer to the continuous choice of projectors and \emph{random} to refer to the choice of the graph. 

This paper presents a cluster of results regarding \emph{generic} quantum satisfiability under the restriction that the projectors $\Pi^m = \ket{\phi}\bra{\phi}$ are rank 1. In Sec.~\ref{sec:product_satisfiability} we introduce and characterize product satisfiability (PRODSAT) on non-random graphs $G$. This essentially classical property is NP-complete, but will allow us to prove that interaction graphs in which each clause (edge) may be matched with a neighboring qubit (node) are quantum satisfiable for \emph{any} choice of projectors. For the marginal case $M=N$, in which these correspond to perfect matchings of the clause-qubit graph, the associated product states are discrete and provide some access to the ground state space, as will be explored in Sec.~\ref{sub:counting_product_states}.  As physicists, we refer to the qubit-clause matchings as \emph{dimer coverings}, although computer scientists will likely prefer to think of perfect matchings and Hall's marriage condition.

The remainder of the paper is especially concerned with mapping out the phase diagram of generic quantum satisfiability on large \emph{random} interaction graphs. As in our previous work\cite{Laumann:2010fk}, we focus on the Erd\"{o}s-Renyi ensemble with clause density $\alpha$, in which each of the potential $\binom{N}{k}$ edges appears with probability $p=\alpha N / \binom{N}{k}$. As one might expect, the ensemble enjoys a satisfiable (SAT) phase at low clause densities $\alpha$ and becomes frustrated (unSAT) at sufficiently high clause density. Originally, we showed the existence of the SAT phase by constructing satisfying product states using something akin to a transfer matrix approach. This algorithm works generically so long as the graph $G$ does not have a hypercore -- that is, a maximal subgraph in which every qubit participates in at least two clauses. This property is guaranteed for $\alpha < \alpha_{hc}(k)$, the critical density beyond which a hypercore appears. These critical values are available in the literature\cite{Molloy:2005lk}; for $k=3$, $\alpha_{hc} \approx 0.81$, while for $k\to\infty$, $\alpha_{hc} \to 0$. 

In Sec.~\ref{sub:application_to_random_graphs}, we apply the characterization of product state satisfiability of Sec.~\ref{sec:product_satisfiability} to the random graph ensemble. This improves the lower bound on the SAT/unSAT transition from $\alpha_{hc}$ to $\alpha_{dc}$, the critical value for the existence of dimer coverings. For $k=3$, $\alpha_{dc} \approx 0.92$ is not much of an improvement over $\alpha_{hc} \approx 0.81$, but as $k \to \infty$, $\alpha_{dc} \to 1$ instead of $0$. Moreover, this is a complete characterization of the phase in which the ensemble is generically satisfiable by product states. If there is a satisfiable phase for larger $\alpha$, its satisfying states will generically be entangled.

At high clause densities, a simple local bound on the generic dimension of the zero energy subspace provided the first evidence for the existence of an unSAT phase. More recently, Bravyi \emph{et al} (BMR) \cite{Bravyi:2009p7817} have dramatically improved this bound by finding the ground state degeneracy of larger clusters (``sunflowers'' and ``nosegay'' graphs) and determining their prevalence within the large random graph. Using these techniques they have placed an upper bound,  $\alpha_c^+$, on the SAT/unSAT transition of $\alpha_c^+ \approx 3.594$ for $k=3$ and with an exponential scaling $\alpha_c^+ \sim 2^k$ for large $k$. 

The gap between the best rigorous lower and upper bounds on the SAT/unSAT transition is distressingly large: $\alpha_c^- = \alpha_{dc} = O(1)$ versus $\alpha_c^+ = O(2^k)$. Moreover, the existence of a satisfiable regime above $\alpha_{dc}$ would provide a large ensemble of large graphs with intrinsically entangled satisfying states. Thus, in Sec.~\ref{sec:beyondPRODSAT}, we address the situation for larger values of $\alpha$ by adducing some
primarily numerical evidence about the existence of random instances which are SAT but not PRODSAT. We note that recently, Ambainis \emph{et al} \cite{Ambainis:2009fv} have proven the existence of such an entangled SAT phase for $k \ge 12$ using their newly proven quantum Lov\'asz local lemma coupled with the characterizations presented in this paper. We will have more to say about this exciting development in the conclusion.


\section{Product satisfiability on general graphs} 
\label{sec:product_satisfiability}

One of the remarkable results in Bravyi's original work on QSAT was the explicit construction of satisfying product states for $k=2$, thanks to which $2$-QSAT turned out to be easy to decide and the concomitant satisfying states easy to construct. 
This poses the question whether there exist regimes for $k\geq3$-QSAT in which analogous results hold. We therefore define:

\begin{defn}
	An instance of QSAT, $(G, \Pi^m)$, is \emph{PRODSAT} if it has a satisfying product state. That is, $\exists \ket{\Psi} = \ket{\psi_1}\otimes\cdots\otimes\ket{\psi_N}$ s.t. $\Pi^m \ket{\Psi} = 0~ \forall m\in G$.
\end{defn}

PRODSAT may be viewed as a decision problem in its own right and, in the
presence of a polynomially small promise gap $\delta$, it is efficiently
verifiable by classical computation. A witness is simply a collection of $2N$
$\CC$ components of a satisfying product state, the energy of which may be
evaluated in linear time. Moreover, by choosing $\Pi_m$ to project onto
computational basis states, it is clear that PRODSAT contains SAT and is
therefore NP-complete.

\begin{figure}[t]
	\centering
		\includegraphics[scale=1]{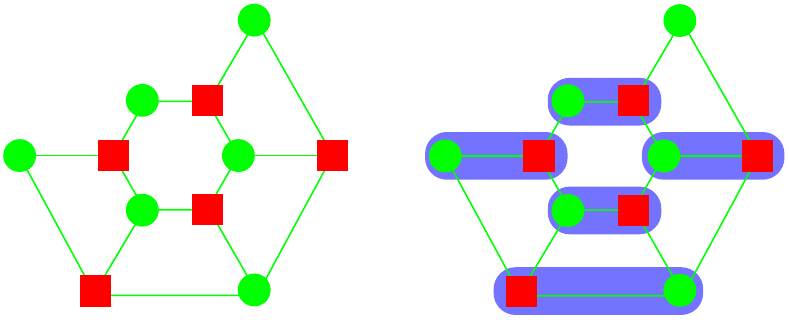}
	\caption{(Color online) (a) Example of a $k=3$ interaction graph with $M<N$. Circles (green) indicate qubits and squares (red) indicate clause projectors that act on adjacent qubits; (b) a dimer (shaded blue) covering that covers all clauses.}
	\label{fig:dimercover}
\end{figure}

In this section, we investigate product satisfiability for fixed graphs $G$
and general choices of rank 1 projectors $\Pi^m = \ket{\phi^m}\bra{\phi^m}$.
We show that $G$ supports satisfying product states on open neighborhoods in
the projector manifold if it has a qubit-clause dimer covering that covers all
of the clauses -- \emph{ie.} an assignment of a unique qubit to each clause
(see Fig.~\ref{fig:dimercover}). Appealing to the algebraic properties of the
space of projectors, this shows that $G$ is in fact PRODSAT for all choices of
$\Pi^m$. It also suggests a counting criterion for the ground state
degeneracy. Thus, we will show

\begin{theorem}\label{thm:prodsat}
	If an interaction graph $G$, viewed as a bipartite factor graph of qubits and clauses, has a dimer covering of its clauses, then
	\begin{enumerate}
		\item there exist open neighborhoods $U \subset (\mathbb{CP}^{2^k-1})^M$ of the rank 1 projector manifold such that $\Pi^m \in U \Rightarrow (G, \Pi^m)$ is PRODSAT.
		\item Moreover, $(G,\Pi^m)$ is PRODSAT for all $\Pi^m \in (\mathbb{CP}^{2^k-1})^M$.
	\end{enumerate}
\end{theorem}

As a practical matter, product satisfiability is a question about the existence of solutions to a collection of $M$ homogeneous equations in $N$ $\CP^1$ degrees of freedom (unless otherwise noted, paired $i$ indices contract, but $m$ indices do not):
\begin{equation}
	\label{eq:prodsat_gen}
	\phi^m_{i_1 i_2\cdots i_k} \psi^{i_1}_{m_1} \psi^{i_2}_{m_2}\cdots \psi^{i_k}_{m_k} = 0~~\forall~ m\in G
\end{equation}
where $\psi^{i_1}_{m_1}$ is the $i_1$'th component of the state on the $m_1$'th qubit and $\phi^{m*}$ is the state onto which $\Pi^m$ projects.
Since we are interested in generic choices of the $\phi^m$, one might expect naive constraint counting to determine whether $(M\le N)$ or not $(M>N)$ a given graph $G$ is generically PRODSAT. Unfortunately, things are not quite so simple because a $G$ with a low average density ($M\le N$) may still have subgraphs that are dense enough to overconstrain the equations.

To make this last observation sharper, let us briefly restrict our attention to product projectors. In this case, Eq.~\eqref{eq:prodsat_gen} reduces to 
\begin{equation}
	\label{eq:prodsat_prodproj}
	 (\phi^{m_1}_{i_1} \psi^{i_1}_{m_1}) (\phi^{m_2}_{i_2} \psi^{i_2}_{m_2}) \cdots (\phi^{m_k}_{i_k} \psi^{i_k}_{m_k}) = 0~~\forall~ m\in G
\end{equation}
This equation is clearly satisfied if any of its factors is satisfied, which requires at least one of the $k$ qubits to be (uniquely) fixed orthogonal to the local projector. Thus, for generic $\phi$ of product form, in which the local projectors on a site $n$ are not parallel, the full set of $M$ equations is solvable if and only if one can uniquely associate a qubit $n$ to each clause $m$. Graphically, this is equivalent to the existence of a dimer covering of the bipartite factor graph of $G$ which covers all $M$ clauses. Note that we refer to matchings as dimer coverings if each clause is
uniquely associated to a qubit even if some qubits are left unpaired. See Fig~\ref{fig:dimercover}.

We now relax the constraint that $\phi$ take product form and argue that the existence of the above dimer covering is still the relevant constraint counting for product satisfiability. 
Let us consider an instance $(G,\phi)$ and assume that it is PRODSAT; \emph{w.l.o.g.} we may choose local bases such that $\ket{\Psi}=\binom{0}{1}\otimes\cdots\otimes\binom{0}{1}$ is a satisfying assignment. Choosing stereographic coordinates $\binom{z_n}{1}$ for each qubit, we find the constraint equation is:
\begin{equation}
	\label{eq:prodsat_stereo}
		\phi^m_{i_1 i_2\cdots i_k} z^{i_1}_{m_1} z^{i_2}_{m_2}\cdots z^{i_k}_{m_k} = 0
\end{equation}
where now superscript $i_1$ indicates exponentiation and the $i$ indices run from $0$ to $1$. Since $z_n = 0$ satisfies this equation, we find that $\phi^m_{00\cdots0} = 0$ necessarily but that the remaining components of the $\phi^m$ are unconstrained. 

Now we perturb $\phi^m \mapsto \phi^m + \delta \phi^m$ and attempt to follow the solution $z_n = 0 \mapsto z_n + \delta z_n$. To linear order, we have (for each clause $m \in G$):
\begin{equation}
	\label{eq:perturb_linear}
	\phi^m_{10\cdots0}\delta z_{m_1} + \phi^m_{01\cdots0} \delta z_{m_2} + \cdots + \phi^m_{00\cdots1}\delta z_{m_k} = - \delta \phi^m_{00\cdots0}
\end{equation}
This is a collection of $M$ sparse linear inhomogeneous equations for the $N$ variables $\delta z_{n}$. Let $\Omega_{m n}$ be the $M\times N$ matrix of coefficients on the left hand side. This matrix connects qubits to clauses and is sparse in the same pattern as the node-edge adjacency matrix $A_{mn}$ of $G$. 

Equation \eqref{eq:perturb_linear} is solvable for arbitrary $\delta \phi^m_{00\dots0}$ if and only if the matrix $\Omega$ is surjective. Trivially, this requires $M \le N$, but it also requires that some $M\times M$ subdeterminant of $\Omega$ be nonzero. Let us define a polynomial discriminant of surjectivity as the tuple of all possible subdeterminants of $\Omega$:
\newcommand{\disc}{\mathrm{disc}}
\begin{equation}
	\label{eq:surj_disc}
	\disc_\tau(\Omega) = \sum_{\sigma\in S_M} (-1)^\sigma \Omega_{1\tau(\sigma(1))} \cdots \Omega_{M \tau(\sigma(M))} 
\end{equation}
where $\tau: M \hookrightarrow N$ runs over all injective maps.
For formal components $\phi$, $\disc_\tau(\Omega)$ is not identically zero if and only if one of the terms in the above sums is not identically zero; each non-zero term corresponds to a particular pairing of clauses and qubits in the adjacency graph which covers all $M$ clauses but need not cover all of the nodes. Thus if $G$ has such a dimer covering, $\Omega$ is almost always surjective, whereas it is never surjective otherwise. 

We can summarize the above analysis as follows:
for $G$ having dimer coverings, generic choices of $\phi$ for which $(G,\phi)$ is PRODSAT may be extended to open neighborhoods around $\phi$ which are PRODSAT. 
Moreover, $G$ has product states at generic (non-parallel) \emph{product} projectors. One can repeat the perturbative analysis of Eq.~\eqref{eq:perturb_linear} near such product projectors and show that the dimer covering condition allows the same extension onto open neighborhoods in the full projector space. 
This proves part 1 of theorem \ref{thm:prodsat}. %
\footnote{For those who prefer a more rigorous looking style, we note that the first order perturbative calculation performed in this section is identical to checking the conditions under which the implicit function theorem provides a smooth map from projector space to the manifold of satisfying product states near a known solution.} %

The above analysis is essentially local, showing that the set $W \subset (\mathbb{CP}^{2^k-1})^M$ of PRODSAT projector choices is nonempty and full dimension; to extend product satisfiability to the \emph{entire} projector manifold, we need to appeal to some basic results in complex projective geometry: 1. Any Zariski-closed full dimension subspace of an irreducible complex projective space (such as $(\mathbb{CP}^{2^k-1})^M$) is in fact the whole space and 2. $W$ is Zariski-closed. This shows part 2 of theorem~\ref{thm:prodsat}. For more details, see Appendix~\ref{app:formal_prodsat_proof}.

Finally, we note that the converse statement also holds (also shown in Appendix~\ref{app:formal_prodsat_proof}):
\begin{theorem}
	\label{thm:notprodsat}
	If $G$ does not have dimer coverings, $(G, \Pi^m)$ is not PRODSAT for almost all $\Pi^m$.
\end{theorem}

\subsection{Counting product states at $M=N$} 
\label{sub:counting_product_states}

We now comment on the marginal case when $M=N$. In this case, one expects the satisfying product states, when they exist, to be discrete. Let us return to projectors of product form and consider a choice $\phi_0$ in which no local projectors on a single site $n$ are parallel, as mentioned above. In this case, it is clear that there is a one-to-one correspondence between dimer coverings $c$ of $G$ and product state solutions $\psi_c$. The number of product states at the special point $\phi_0$ is given by the number of dimer coverings.

Now, one can pick any of the $\psi_c$ and repeat the local perturbative analysis around $\phi_0$. This will show a unique way to continuously extend $\psi_c$ in an open neighborhood of $\phi_0$. Thus, at least on open neighborhoods within projector space, the number of product states is bounded below by the number of dimer coverings. Since the problematic points in $\phi$ space for this local analysis are actually degenerate points with higher satisfiability, this should be a lower bound on the number of product states for any $\phi$ and it should be the exact counting almost everywhere.

With a handle on all of these product states, one might hope to make some sharp statements about the QSAT dimension of $G$ -- that is, the generic $R_G = |\ker(H)|$ for the graph $G$. Let $PS = \mathrm{Span}\set{\psi_c}$ be the span of the dimer covering states at some generic product projector point $\phi_0$. If $R_{PS} = |PS|$ of the $\psi_c$ are linearly independent, they will remain so on an open neighborhood of $\phi_0$, since the determinant of their overlap matrix is a smooth function. We recall that the QSAT dimension $R_G$ takes its minimal value \emph{almost everywhere} on projector space \cite{Laumann:2010fk}. Since $R_{PS}$ lower bounds $R_G$ on an open neighborhood, it must therefore lower bound $R_G$ for all projectors $\phi$.

Using the dimer covering states to characterize the full SAT subspace $\ker(H)$ for a given graph $G$ requires answering two questions: 1. Are the product states linearly independent? and 2. Do they span $\ker(H)$? For some simple families of graphs, it is possible to prove the linear independence of the dimer states by exploiting the geometry of their dimer coverings.  On the other hand, it is easy to construct graphs where the dimer states are not independent: for a fully connected bipartite factor graph, the number of dimer coverings is $N!$, which is significantly greater than the size $2^N$ of the Hilbert space. 

The second question is harder to address analytically. Product states do span the kernel for many graphs that we have studied numerically, but we also know examples of small graphs for which they do not. Rather, $R_G$ can be strictly greater than the number of dimer coverings. Indeed, $H$ restricted to product projectors may have an even greater satisfying dimension than it does for general entangled projectors, even though the number of product states should be no greater.



\section{Application to random graphs} 
\label{sub:application_to_random_graphs}

From the previous section, we know that the existence of dimer coverings on $G$ implies quantum satisfiability -- and that their nonexistence shows satisfying product states are no longer generic. For the random graph ensemble, we can therefore lower bound the QSAT transition $\alpha_c$ by the threshold for the presence of dimer coverings $\alpha_{dc}$. This turns out to be precisely the point at which the hypercore of the random graph reaches the critical density $M_{hc} = N_{hc}$, beyond which it is certainly impossible to dimer cover the clauses. In general, the hypercore emerges at $\alpha_{hc}<1$ with a clause density less than 1 and as $\alpha$ grows, it gets denser. Only at $\alpha_{dc} > \alpha_{hc}$ does it become critically constrained. See Table~\ref{tab:alphacrit} for numeric values of $\alpha_{dc}$. 

In order to show that dimer coverings exist below $\alpha_{dc}$, we rely on a somewhat indirect argument based on the results of Mezard \emph{et al.}\cite{Mezard:2003p5977} regarding the random $k$-XORSAT problem, or, in physics parlance, the $p$-spin Ising glass (with $p=k$). In this problem, $N$ classical bits live on an interaction graph $G$ whose (hyper-)edges define parity check constraints. That is, for each edge $m$, we must choose $J_m \in \set{0, 1}$ and then determine whether the following linear constraint equations are satisfiable:
\begin{equation}
	\label{eq:xorsat}
	x_{m_1} + x_{m_2} + \cdots + x_{m_k} = J_m
\end{equation}
where $x_n$ are the bits and all arithmetic takes place in the field $\ZZ_2$. 

This defines a linear system of inhomogeneous equations of exactly the same form as what we discovered when linearizing PRODSAT:
\begin{equation}
	A_{mn} x_n = J_m
\end{equation}
where $A$ is the node-edge adjacency matrix of $G$. In this case, $A$ is surjective if and only if all choices of $J_m$ result in satisfiable instances of XORSAT. Moreover, if $A$ is surjective over $\ZZ_2$, so is the sparse matrix $\Omega_{mn} = \phi_{mn} A_{mn}$ over $\CC$ with arbitrary components $\phi_{mn}$: if $\Omega$ were not generically surjective, the rank discriminating polynomials $\disc_\tau(\Omega)$ of Eq.~\ref{eq:surj_disc} would be identically zero, including when mapped homomorphically to $\ZZ_2$. 

Mezard \emph{et al} show that the random XORSAT problem is unfrustrated for $\alpha < \alpha_{dc}$. Indeed, they show precisely that with probability going to 1 in the thermodynamic limit, the random graph $G$ is XORSAT for \emph{all} choices of $J_m$. In particular, this implies that $A$ is surjective for $\alpha < \alpha_{dc}$ and therefore that $\Omega$ is as well, dimer coverings exist and the QSAT problem is satisfiable. Meanwhile, for $\alpha > \alpha_{dc}$, $M>N$ on the hypercore and the dimer coverings are gone.

\begin{figure}[t]
	\centering
		\includegraphics[width=0.8\linewidth]{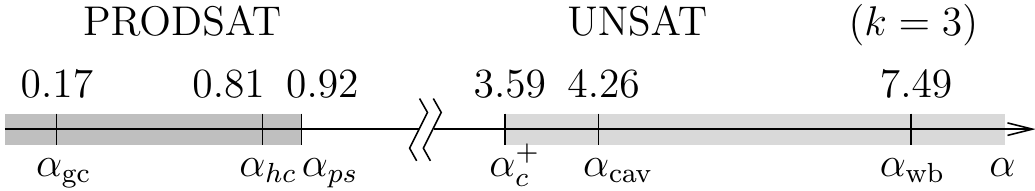}
	\caption{Phase diagram of random $k$-QSAT for $k=3$. The labeled transitions are (left to right): the emergence of the giant component ($\alpha_{gc}$); the emergence of the hypercore ($\alpha_{hc}$); the disappearance of product states ($\alpha_{ps}=\alpha_{dc}$); the best current upper bound on satisfiability due to Bravyi \emph{et al}\cite{Bravyi:2009p7817} ($\alpha^+_{c}$); the classical cavity transition for $k$-SAT ($\alpha_{cav}$) and the weak Pauling bound ($\alpha_{wb}$). At large $k$, the known upper bounds all scale exponentially while the best lower bound ($\alpha_{ps}$) is bounded above by $\alpha = 1$.}
	\label{fig:pd-3-qsat}
\end{figure}

We note that the previously known lower bound was given by $\alpha_{hc}$, the threshold for the emergence of a hypercore in $G$. As $k$ becomes large, $\alpha_{hc}$ approaches zero while $\alpha_{dc}$ approaches 1 (from below). This significantly raises the known lower bound for quantum satisfiability, but it still does not approach the best known upper bounds, which scale with $2^k$. See the phase diagram in Fig.~\ref{fig:pd-3-qsat} for the numerical values at $k=3$.

\begin{table}[t]
	\begin{center}
	\begin{tabular}{c| c c c | c c |}
		\cline{2-6}
		& \multicolumn{3}{|c|}{Quantum} & \multicolumn{2}{|c|}{Classical} \\
		\hline
		\multicolumn{1}{|c|}{$k$} & $\alpha_{hc}$ & $\alpha_{ps}$ & $\alpha^+_c$ & $\alpha^-_{cl}$ & $\alpha^+_{cl}$ \\
		\hline
		\multicolumn{1}{|c|}{2} & 0.5 & 0.5 & 0.5 & 1 & 1 \\
		\multicolumn{1}{|c|}{3} & 0.82 & 0.92 & 3.59 & 3.52 & 4.49 \\
		\multicolumn{1}{|c|}{4} & 0.77 & 0.98 & 7.98 & 7.91 & 10.23 \\
		\multicolumn{1}{|c|}{5} & 0.70 & 0.99 & 16.00 & 18.79 & 21.33 \\ 
		\hline
		\multicolumn{1}{|c|}{$\infty$} & $0^+$ & $1^-$ & $2^{k-1} \ln 2$ & $2^k \ln 2$ & $2^k \ln 2 $ \\
		\hline
	\end{tabular}
	\end{center}
	\caption{Summary of critical values for various random $k$-(Q)SAT properties. The values for the emergence of a hypercore $\alpha_{hc}$ and for the dimer coverability/product satisfiability $\alpha_{ps} = \alpha_{dc}$ of the random interaction graph may be found in \cite{Mezard:2003p5977}. $\alpha^+_c(3)$ comes from the BMR `nosegay' bound\cite{Bravyi:2009p7817} while the values at larger $k$ and in the large $k$ limit come from the `sunflower' bound (cf. Appendix~\ref{app:eval_sunflower}). For comparison, we include the best known rigorous bounds on the classical satisfibility transition \cite{Ach09HBSAT,Diaz:2008eu}.}
	\label{tab:alphacrit}
\end{table}


\section{Beyond PRODSAT} 
\label{sec:beyondPRODSAT}

In the previous section, we argued that random $k$-QSAT is PRODSAT for all $k$, at clause densities $\alpha < \alpha_{dc}(k) = \alpha_{ps} (k)$ with $\alpha_{ps} (2) = \frac{1}{2}, ~\alpha_{ps} (3) = .92 \dots$ and $\alpha_{ps} (k \to \infty) \to 1^-$. (A heuristic way of understanding the latter limit is provided by the observation that for large $k$, the probability of having sites with low coordination is suppressed exponentially $\sim e^{-\alpha k}$, so that the density of the random graph and its 2-core are almost the same.)

What happens for $\alpha > \alpha_{ps} (k)$? Reliable information on this only exists for $k = 2$: as the ensemble goes unPRODSAT, it immediately goes unSAT as well. For $k \geq 3$, no such result has as yet been established. Rather, the best upper bounds on the extent of the SAT phase, due to BMR\cite{Bravyi:2009p7817}, are well separated from the PRODSAT transition: a  detailed analysis of the equations they derive yields the upper bounds of Table~\ref{tab:alphacrit}. The large-$k$ asymptote of this bound is $\alpha^+_c (k) = 2^{k-1} \ln 2 + O \left( \frac{1}{\ln k} \right)$, which is {\em exponential} in $k$, leaving plenty of space above $\alpha_{ps} = 1^-$ for a SAT--unPRODSAT phase.

In the following, we adduce some numerical pointers regarding the existence of such a phase, which is arguably among the most interesting open questions in this problem.

\subsection{Direct search for SAT--unPRODSAT instances}
\label{sec:SATunPRODSAT}

\begin{figure}[t]
	\centering
		\includegraphics[width=0.9\linewidth]{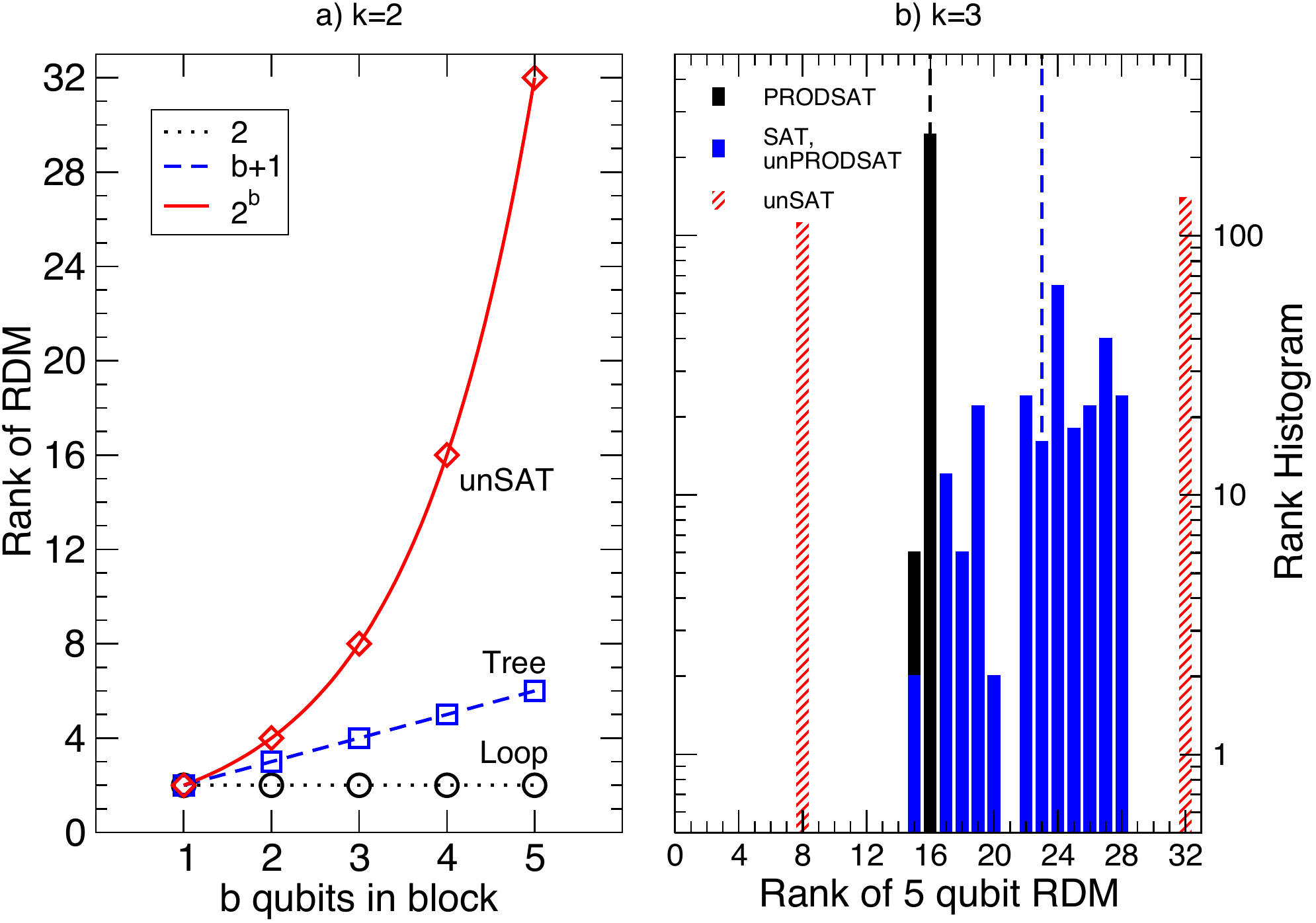}
\caption{(Color online) Rank analysis of reduced density matrices (RDM) in a ground state: (a) \underline{$k=2$}: Rank of the reduced density matrix containing $b=1,\ldots,5$ qubits for different types of 
interaction graphs.  (b) \underline{$k=3$}: Histogram of ranks of $b=5$ reduced density matrices for all 252 different partitions ($N=M=10$) for three different instances.
}
\label{fig:rdmranks-k_3-k_4}
\end{figure}

First, we have directly searched for instances -- irrespective of their probability in the random graph ensemble -- which are SAT--unPRODSAT.  Our search for satisfying assignments proceeded via (i) exact diagonalisation of systems with up to 14 qubits and (ii) a numerical nonlinear equation solver for finding product solutions. Our results were all consistent with the picture of product states outlined above. In addition, we found graphs with  satisfying assignments with $M>N$. In these cases method (ii) failed to turn up any product ground states. We also diagnosed directly the presence of non-product ground states by considering the rank of the reduced density matrix corresponding to a ground state after an arbitrary portion of the system is traced over, \cf
Fig.~\ref{fig:rdmranks-k_3-k_4}. The resulting rank of the density matrix is bounded above by the number of zero-energy states, provided these are spanned by product states 
(c.f. Appendix~\ref{app:rdm_diagnostics} for a proof). As an illustration, we show the corresponding rank for the 2-SAT problem in Fig.~\ref{fig:rdmranks-k_3-k_4}(a) for reduced density matrices of up to five qubits from instances containing 13 qubits. While the SAT instances show the expected rank for product states, the unSAT instance has full rank, suggesting generically entangled wavefunctions in the $k=2$ unSAT phase. In Fig.~\ref{fig:rdmranks-k_3-k_4}(b) we performed a similar analysis for $k=3$. We determined rank histograms for all 252 five-qubit
reduced density matrices of $N=M=10$ systems for the three different instances  listed in~\footnote{
The three $N$=$M$=$10$ graph instances are listed here for convenience: 
a) PRODSAT:
	\{\{3,5,8\}, \{5,6,7\}, \{0,6,8\}, \{1,3,5\}, \{0,2,5\}, \{1,3,9\}, \{1,4,9\}, \{0,1,5\}, \{2,3,7\}, \{0,1,2\}\}
b) SAT-unPRODSAT:
	\{\{0,1,5\}, \{0,4,5\}, \{2,5,8\}, \{5,7,9\}, \{5,6,7\}, \{5,7,8\}, \{6,8,9\}, \{0,3,5\}, \{4,6,8\}, \{2,4,9\}\}
c) unSAT:
	\{\{2,7,8\}, \{0,4,7\}, \{4,5,6\}, \{1,5,6\}, \{1,6,9\}, \{3,5,7\},  \{0,3,7\}, \{5,6,7\}, \{1,3,7\}, \{1,6,7\}\}
}.
The PRODSAT instance leads to a histogram where the rank is bounded by the number of satisfying ground states (16 for this instance). The SAT--unPRODSAT instance shows ranks which are both smaller and larger than the number of satisfying ground states (23 for this instance), but do not reach the maximum of $2^5=32$. Finally, the unSAT instance has full rank $32$ on the two-core (the rank 8 occurring for some partitions is due to a satisfied dangling clause). We do indeed find that, in the presence of $M > N$ on a (sub)graph, entangled states necessarily appear in any basis of the ground states. 

\subsection{Finite size scaling}
\label{sec:larger_k}

\begin{figure}[t]
	\centering
		\includegraphics[width=0.9\linewidth]{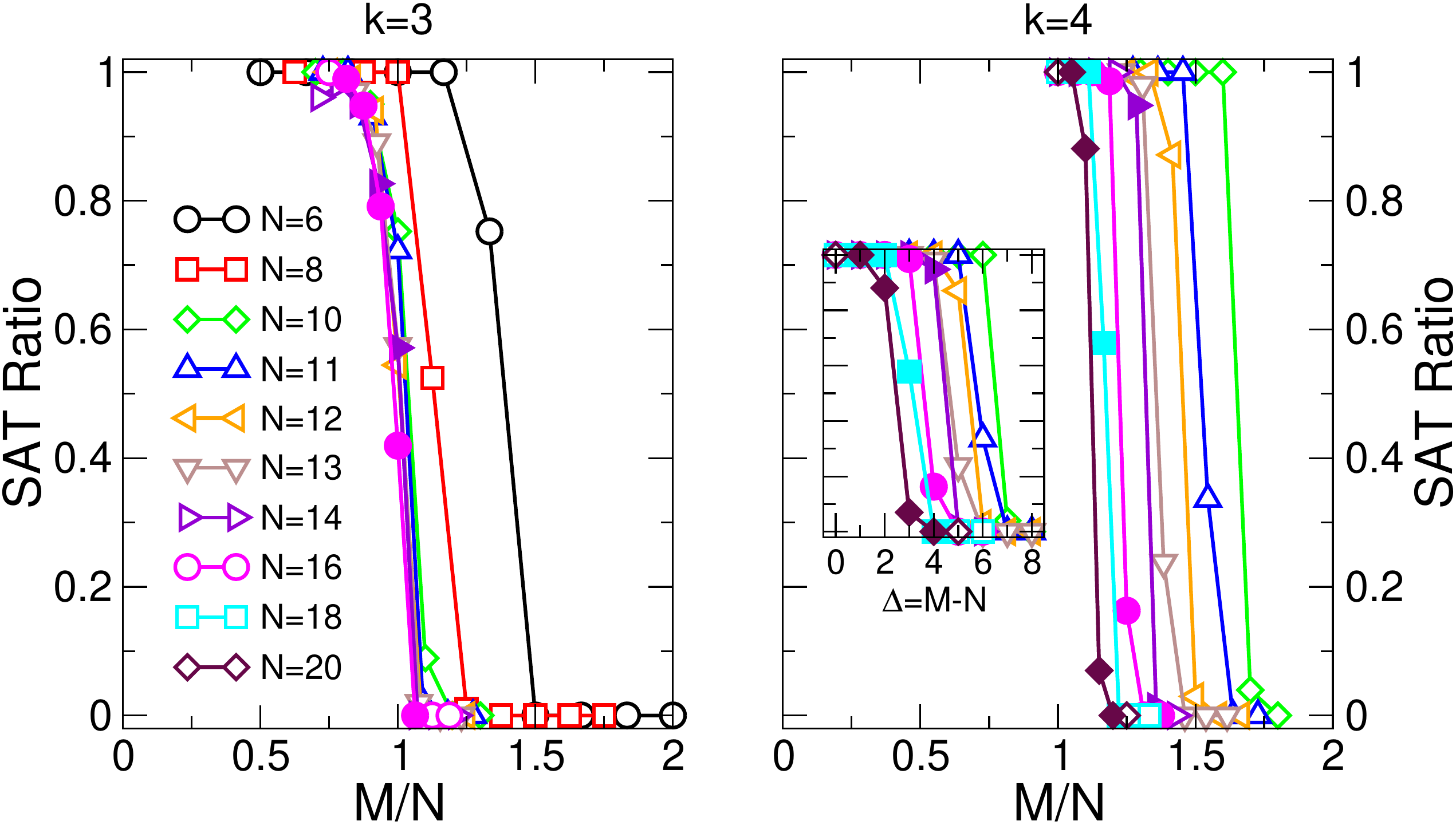}
	\caption{(Color online) Probability to find SAT instances among 101 random graphs for (a) $k=3$ and (b) $k=4$. Numerical results obtained by complete diagonalization for $N\le13$ and
	iterative diagonalization for $N=14,16$ $(k=3,4)$ and $18,20$ ($k=4$). For the latter system sizes filled symbols indicate the occurrence of (numerically) undecidable instances.}
	\label{fig:satratio-k_3-k_4}
\end{figure}

\begin{figure}[t]
	\centering
		\includegraphics[width=0.8\linewidth]{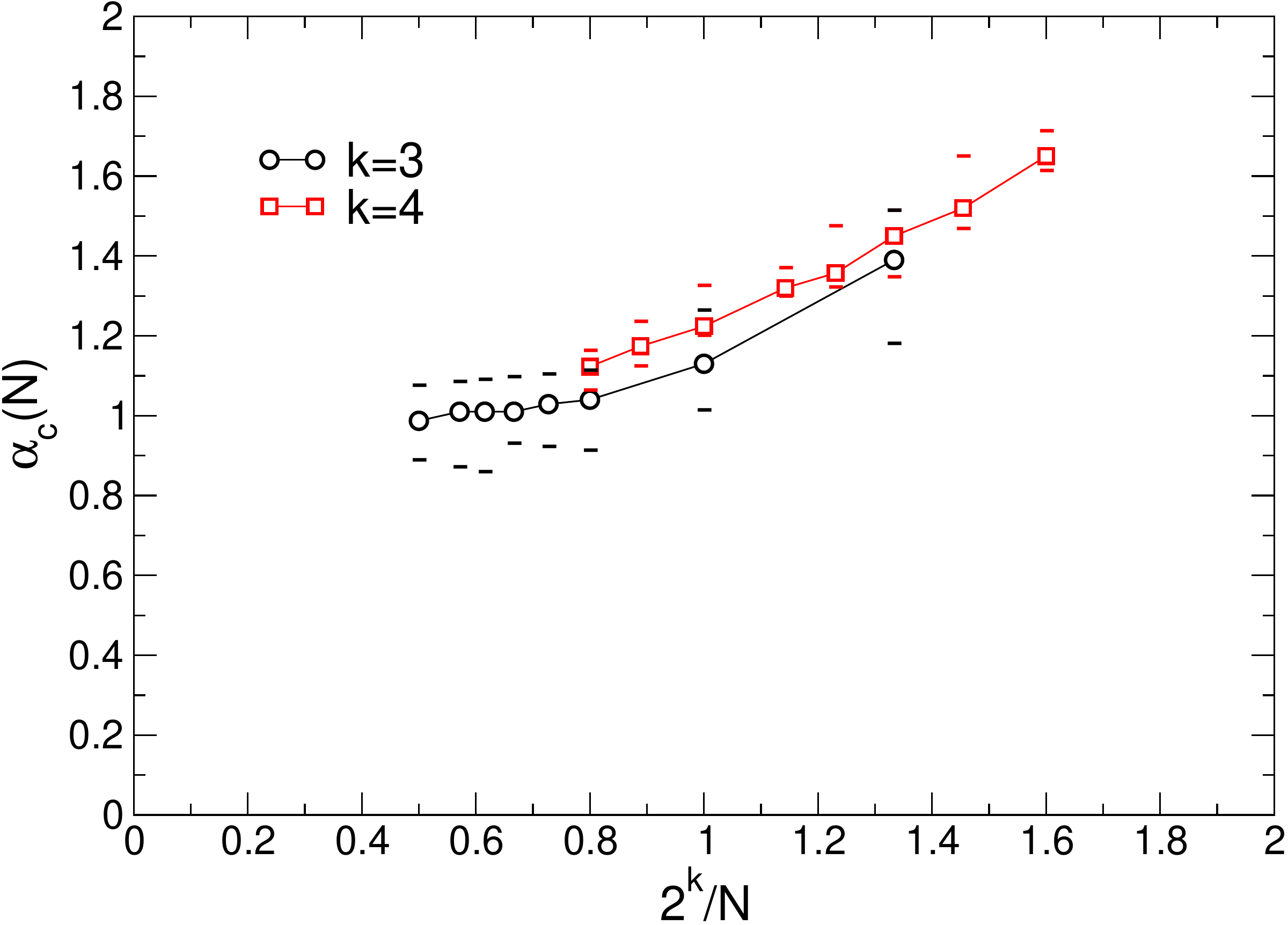}
	\caption{(Color online) Finite size scaling of $\alpha_c(N)$ for $k=3,4$ obtained from the data displayed in Fig.~\protect{\ref{fig:satratio-k_3-k_4}}. 
	The empty symbols denote $\alpha$ values where the SAT ratio is 0.5. The lower (upper) ticks 
	correspond to the largest $\alpha$ values with SAT ratio~$>0.9$ (smallest $\alpha$ values with SAT ratio~$<0.1$).}
	\label{fig:criticalalpha-k_3-k_4}
\end{figure}

Where is the the SAT/unSAT transition $\alpha_c$? Numerical studies on finite-site systems offer limited information as the properties of small random graphs are quite different from those of large ones. 
We do consistently find the presence of SAT instances with $N > M$: for 3-SAT, we investigated 101 random graphs for sizes with up to 13 sites using 
complete diagonalization and up to 20 sites using an iterative diagonalization technique, see left panel of Fig.~\ref{fig:satratio-k_3-k_4}.
While complete diagonalization is limited by the growing matrix size but otherwise numerically stable, the iterative Lanczos procedure faces challenges when resolving small energy gaps occurring at the SAT-unSAT transition. This leads to some instances becoming undecidable, i.e. not converging within several thousand iterations. The presence of undecidable instances is
denoted by filled symbols~\footnote{Among the 101 graphs we found the following maximum number of undecidable instances: 
$N=14, k=3,$ max 4 undecided;
$N=16, k=3,$ max 9 undecidable;
$N=14, k=4,$ max 24 undecidable;
$N=16, k=4,$ max 57 undecidable;
$N=18, k=4,$ max 62 undecidable;
$N=20, k=4,$ max 58 undecidable.
}.
For 3-SAT, we find that curves for larger $N$ nicely converge, suggesting $\alpha_c \approx 1 \pm 0.06$, where the error bar is estimated from the largest $N$ in Fig.~\ref{fig:criticalalpha-k_3-k_4}.
For 4-SAT, instances are SAT for larger values of $\Delta =M-N$ (for fixed $N$), even though $\Delta$ appears to decrease with $N$ (inset in right panel of Fig.~\ref{fig:satratio-k_3-k_4}).
The finite size scaling shown in Fig.~\ref{fig:criticalalpha-k_3-k_4} is also compatible with $\alpha_c \approx 1$, but with a larger uncertainty of $\approx 0.2$.


Thus, we find many examples of generically SAT--unPRODSAT graphs $G$ and even some evidence that the unSAT phase transition in the random ensemble lies at $\alpha > \alpha_{ps}$.


\section{In closing} 
\label{sec:conclusion}

In this paper we have reported considerable progress in understanding product satisfiability. Applied to the random graph ensembles introduced in \cite{Laumann:2010fk}, it yields a definitive lower limit to the clause density at which satisfiability necessarily requires entangled states and certainly an improved lower bound to the location of the SAT-unSAT phase transition. Between the work reported in this paper and by BMR \cite{Bravyi:2009p7817}, there has thus been a fair amount of progress in narrowing the location of this phase transition from the initial work \cite{Laumann:2010fk}. 

Nonetheless, there remains a large gap between the  best lower bound $\alpha^-_c = \alpha_{ps} = O(1)$ and the best upper bound $\alpha^+_c = O(2^k)$.  By contrast, the classical random $k$-SAT problem has tight bounds on its satisfiability transition that scale as $2^k$ \cite{Ach09HBSAT}. It is not obvious how to translate the techniques used in proving the tight lower bounds for the classical problem to the quantum case.

Ambainis and co-workers\cite{Ambainis:2009fv} had recently proven a quantum
version of the Lov\'{a}sz local lemma, which directly shows the quantum
satisfiability of graphs $G$ of bounded degree. Due to this degree
restriction, the lemma does not directly apply to the Erd\"os random graph
ensemble. Thus, they have used the characterization of product satisfiability
shown here coupled with the quantum Lov\'asz lemma in order to show that a
wide SAT-unPRODSAT phase in the parameter $\alpha$ exists: the SAT regime
extends to $\alpha_c > 2^k/(12 \cdot e \cdot k^2)$ which, for $k \ge 13$, is
significantly beyond the PRODSAT transition at $\alpha_{ps} \approx 1$.

Many other questions remain open and it is to these that we turn in closing:

\begin{enumerate}
	\item Does the SAT-unPRODSAT phase exist for all $k \ge 3$ or does it really only arise at some larger $k$?
	
	\item Does an easily checked classical graph property encode generic quantum satisfiability? Does a straightforward function compute the generic dimension $R_G$?
	
	\item What can be said about higher rank projectors? The product satisfiability condition translates straightforwardly to the higher rank case by simply treating a rank $r$ projector as a stack of rank $1$ projectors. Thus, higher rank QSAT instances may naturally yield more entangled ground states than the rank 1 case. Indeed, preliminary work indicates that this is the case\cite{Movassagh:2010fk}.
	
\end{enumerate}


\section*{Acknowledgements} 
\label{sec:acknowledgements}

We thank S. Bravyi, C. Moore, T. Osborne, D. Nagaj, A. Obus and especially O. Sattath for many enlightening conversations. We would also like to thank the generous hospitality of the Erwin Schr\"odinger Institute in Vienna where parts of this work were completed. 


\appendix

\section{Formal PRODSAT proof} 
\label{app:formal_prodsat_proof}

We expand on the complex algebraic details of theorems \ref{thm:prodsat} and \ref{thm:notprodsat}. That is, we show that interaction graphs $G$ with dimer coverings are PRODSAT \emph{for any} choice of projectors $\{\Pi^m\}_{m=1}^M$ and conversely that $G$ without dimer coverings are not PRODSAT $\emph{for almost all}$ such choices. The relevant algebraic geometry background can be found in \cite{Milne:2009kx,Mumford:1995uq}.

\begin{proof}
	
Let $P = (\mathbb{CP}^{2^k-1})^M$ be the space of $k$-qubit projectors, $S =
(\mathbb{CP}^1)^N$ be the space of product states, and $V \subset P\times S$
be the space of pairs of $(\{\Pi^m\}_{m=1}^M,\{\ket{\psi_n}\}_{n=1}^N)$ such
that $\ket{\Psi} = \ket{\psi_1} \otimes \cdots \otimes \ket{\psi_N}$ satisfies
the projectors $\Pi^m$. We note that $P$, $S$ and $P\times S$ are each
irreducible complex projective spaces in the Zariski topology and that $V$, as
the zero-locus of collection of homogeneous algebraic constraints, is a closed
subspace of $P\times S$. Let $\pi: P\times S \rightarrow P$ be the projection
map onto the first component. Since $S$ is projective, this map is closed.

We are interested in the set of projectors $W \subseteq P$ which are
satisfiable by product states. These are precisely the projectors in the image
of $V$ under the projection $\pi$: 
\begin{equation} 
	W = \pi(V) 
\end{equation}
Since $\pi$ is closed, $W$ is closed. By the local arguments of
Section~\ref{sec:product_satisfiability}, we know that $W$ is full dimension
and nonempty. Since $P$ is irreducible and projective, it has no proper full dimension closed subspaces; \emph{ie.} $W = P$.

Conversely, if $G$ does not have dimer coverings, there exist choices of product projectors $\Pi^m$ which are clearly unsatisfiable by product states (see Eq.~\eqref{eq:prodsat_prodproj}). Hence $W$ is a closed proper subspace of $P$ and cannot be full dimension. Thus, $W$ has codimension at least 1 in $P$ and almost all choices of projector $\Pi^m$ will fail to be product satisfiability.
\end{proof}


\section{Reduced density matrix diagnostic for product states} 
\label{app:rdm_diagnostics}

Let us consider a graph $\Lambda$  and a wavefunction $| \Psi \rangle$, which is
expressible as a sum of wavefunctions each of which is a product wavefunction on
the sites $i \in \Lambda$. Let $\gamma \subset \Lambda$, and let $\rho =
| \Psi \rangle \langle \Psi |$ and $\tilde{\rho}_\gamma =
\mathbf{Tr}_{\Lambda \setminus \gamma} \rho$.
Then $\text{rank}~
\tilde{\rho}_\gamma \leq N^{(\gamma)}_0 \leq N_0$ where $N_0$ is the number of
linearly independent product states, and $N^{(\gamma)}_0$ the dimension of their span when
restricted to $\gamma$.

Let us label a product basis for the ground state manifold by: $\left\{
  \Phi_\alpha \left| \alpha = 1 \dots \right. N_0 \right\}$ with $|
\Phi_\alpha \rangle = \otimes^{n_s}_{i = 1} | \phi^\alpha_i \rangle$,
$\left< \phi^\alpha_i | \phi^\alpha_i \right> = 1$. Here $||\Lambda||=n_s$.
Thus, $\left<
  \Phi_\alpha | \Phi_\alpha \right> = 1$ (normalised) and $\left<
  \Phi_\alpha | \Phi_\beta \right> = \prod^{n_s}_{i=1} \left<
  \phi^\alpha_i | \phi^\beta_i \right>$ (not necessarily orthogonal).

A general ground state is
\begin{equation}
| \Psi \rangle = \frac{1}{\aleph} \sum^{N_0}_{\alpha=1} \lambda_\alpha
|\Phi_\alpha \rangle
\label{eq:01}
\end{equation}

\noindent with
\begin{equation}
\aleph^2 = \sum^{N_0}_{\alpha,\beta=1} \lambda_\alpha \lambda^*_\beta
\prod^{n_s}_{i=1} \left< \phi^i_\beta | \phi^i_\alpha \right>~~~,
\label{eq:02}
\end{equation}

\noindent for which
\begin{equation}
\rho = | \Psi \rangle \langle \Psi | = \frac{1}{\aleph^2}
\sum^{N_0}_{\alpha,\beta=1} \lambda^*_\beta \lambda_\alpha | \Phi_\alpha \rangle \langle \Phi_\beta |~~~.
\label{eq:03}
\end{equation}

Tracing over the qubits on subset $\Lambda\setminus\gamma$ of the set $\Lambda$ yields:
\begin{widetext}
\begin{equation}
\tilde{\rho}_\gamma = \frac{1}{\aleph^2} \sum^{N_0}_{\alpha,\beta=1}
\lambda^*_\beta \lambda_\alpha \left[ \otimes_{i' \in \gamma}
|\phi^\alpha_{i'} \rangle \right] \left[ \otimes_{j' \in \gamma}
\langle \phi^\beta_{j'} | \right] \prod_{k' \in \Lambda \setminus
\gamma} \left< \phi^\beta_{k'} | \phi^\alpha_{k'} \right> ~~~.
\label{eq:04}
\end{equation}
$\tilde{\rho}_\gamma$ is at most an $N_0 \times N_0$
matrix, with at most rank $N_0$ irrespective of
the choice of $\gamma$.

If $\text{dim}\left( \text{span} \left\{ \otimes_{i' \in \gamma} |
    \phi^\alpha_{i'} \rangle | \alpha = 1 \dots N_0 \right\} \right) =
N^{(\gamma)}_0 < N_0$, then we can write, for $\alpha = N_0$ (say): $\otimes_{i'
  \in \gamma} | \phi^{N_0}_{i'}\rangle = \sum^{N_0 - 1}_{\alpha =
  1} \mu^{(N_0)}_\alpha \left[ \otimes_{i' \in \gamma} | \phi^\alpha_{i'}
  \rangle \right]$, so that terms like
\begin{equation}
\left[ \otimes_{i' \epsilon \gamma} |\phi^{N_0}_{i'} \rangle \right]
\left[ \otimes_{j' \epsilon \gamma} \langle \phi^{N_0}_{j'} | \right] =
\sum^{N_0-1}_{\alpha,\beta=1} \mu^{(N_0)}_\alpha \mu^{*(N_0)}_\beta \left[ \otimes_{i'
  \epsilon \gamma} |\phi^\alpha_{i'} \rangle \right] \left[
\otimes_{j' \in \gamma} \langle \phi^\beta_{j'} | \right]~~~,
\label{eq:05}
\end{equation}

\end{widetext}
take on the same form as the ones in (\ref{eq:04}). Using such substitutions repeatedly,
one finds that the reduced dimensionality manifestly has a rank
of at most $N_0^{(\gamma)}$.

In particular, for the case of 2-QSAT and $\Lambda$ being a tree $(|| \Lambda || + 1 =
N_0)$, $N^{(\gamma)}_0 = || \gamma || + 1$. When a qubit being
traced out is a dangling one, this result is obvious. In the case of an internal qubit (labelled 2, say),
we note that the Bravyi construction on
$\Lambda$ induces one on $\gamma$ via $\top^{(\gamma)}_{13} = \top_{12}
\top_{23}$, where $\top_{ij}$ represents the Bravyi transfer matrix
between for the projector ${ij}$,
so that the resulting state counting is that of the smaller graph.


\section{Improved estimate of $\alpha_c^+$ from BMR sunflowers} 
\label{app:eval_sunflower}

In this section we provide an integral representation of the upper bound on the ground state dimension obtained in \cite{Bravyi:2009p7817} by studying `sunflower' graphs for generic $k$. By means of this we can obtain upper bounds $\alpha_c^+(k)$ for $k\geq 4$ and an asymptotic estimate for large $k$.
From \cite{Bravyi:2009p7817} we can read an upper bound to the dimension of the satisfying space $R_G$ as
\begin{widetext}
\begin{equation}
\lim_{n\to\infty}\frac{1}{n}R_G\leq \ln 2 +\sum_{d=0}^\infty a_d\left(d\ \ln (1-2^{-k+1})+\ln\left(1+\frac{d}{2^k-2}\right)\right)\equiv S,
\end{equation}
where
\begin{equation}
a_d=\int_0^1 dt\frac{(k\alpha t^{k-1})^d}{d!}e^{-k\alpha t^{k-1}}\equiv\int_0^1 dt\ a_d(t).
\end{equation}
By inverting the sum over $d$ and the integral over $t$ we find
\begin{equation}
S=\ln 2+\alpha\ln(1-2^{-k+1})+\int_0^1dt\langle\ln\left(1+\frac{d}{2^k-2}\right)\rangle_{t}.
\end{equation}
The last is a $t$-dependent average with the Poisson distribution $a_d(t)$. We use the representation
\begin{equation}
\ln(1+x)=\int_0^\infty ds\frac{e^{-s}}{s}(1-e^{-s x})
\end{equation}
to perform the average over $d$ first (using the generating function of the Poisson distribution) and the integral over $t$ afterwards. We are left with
\begin{equation}
S=\ln 2+\alpha\ln(1-2^{-k+1})+\int_0^\infty ds\frac{e^{-s}}{s}\left(1-\frac{q^{1/(k-1)}}{k-1}\left(\Gamma\left(\frac{1}{k-1}\right)-\Gamma\left(\frac{1}{k-1},q\right)\right)\right)
\end{equation}
where $q=k\alpha(1-e^{-s/(2^k-2)})$ and $\Gamma(z,x)$ is the incomplete gamma function (with $\Gamma(z,0)=\Gamma(z)$). This integral representation, although seemingly complicated, is perfectly suited for numerical integration, since the integrand is always finite and cut-off at $s=O(1)$, beyond which it decreases exponentially in $s$.
\end{widetext}

One can obtain the numbers quoted in the text for the threshold $a_c^+(k)$ to arbitrary accuracy using Mathematica. Moreover in the limit $k,\alpha\to\infty$ and $\alpha/2^k$ finite, relevant for the large $k$ asymptotics, we can bound the value of the integral, solve for $S=0$ and find the quoted result
\begin{equation}
\alpha_c^+=2^k\frac{\ln2}{2}+O(1/\ln k).
\label{eq:aclargek}
\end{equation}
it is worth noting that this bound is half of the value of the \emph{lower} bound of $k$-SAT which is known to become exact in the large $k$ limit.

We have not been able to analyze in the same detail the upper bound given by the `nosegay' graphs of  \cite{Bravyi:2009p7817}. Although this bound would be stricter for \emph{any} $k$ than that coming from sunflower graphs, our preliminary results point towards the same asymptotic scaling (\ref{eq:aclargek}). It would be interesting to understand whether for large $k$ this bound is tight, as happens in classical random $k$-SAT.


\bibliographystyle{apsrev-nourl}
\bibliography{qsat}

\end{document}